\documentclass[runningheads]{llncs}
\usepackage{graphicx}
\usepackage{hyperref}
\usepackage{marvosym}
\usepackage{cite}
\usepackage{amsmath}
\usepackage{amssymb}
\usepackage{pfnote}
\usepackage{multicol}
\usepackage{multirow}
\usepackage{booktabs}
\usepackage{float}
\begin{document}
\title{FedAutoMRI: Federated Neural Architecture Search for MR Image Reconstruction}
\titlerunning{FedNAS for MRI Reconstruction}
%
\author{Ruoyou Wu\inst{1,2,3} 
\and Cheng Li\inst{1} 
\and Juan Zou\inst{1,4} 
\and Shanshan Wang\inst{1,2}$^{\href{mailto: sswang@siat.ac.cn}{(\textrm{\Letter})}}$
}
\authorrunning{R.Wu et al.}
%
\institute{Paul C. Lauterbur Research Center for Biomedical Imaging, Shenzhen Institute of Advanced Technology, Chinese Academy of Sciences, Shenzhen 518055, China 
\email{ss.wang@siat.ac.cn}\\
\and Peng Cheng Laboratory, Shenzhen 518055, China
\and University of Chinese Academy of Sciences, Beijing 100049, China
\and School of Physics and Optoelectronics, Xiangtan University, Xiangtan 411105, China
}
%
\maketitle              
\begin{abstract}
Centralized training methods have shown promising results in MR image reconstruction, but privacy concerns arise when gathering data from multiple institutions. Federated learning, a distributed collaborative training scheme, can utilize multi-center data without the need to transfer data between institutions. However, existing federated learning MR image reconstruction methods rely on manually designed models which have extensive parameters and suffer from performance degradation when facing heterogeneous data distributions. To this end, this paper proposes a novel FederAted neUral archiTecture search approach fOr MR Image reconstruction (FedAutoMRI). The proposed method utilizes differentiable architecture search to automatically find the optimal network architecture. In addition, an exponential moving average method is introduced to improve the robustness of the client model to address the data heterogeneity issue. To the best of our knowledge, this is the first work to use federated neural architecture search for MR image reconstruction. Experimental results demonstrate that our proposed FedAutoMRI can achieve promising performances while utilizing a lightweight model with only a small number of model parameters compared to the classical federated learning methods.

\end{abstract}
\section{Introduction}
Magnetic resonance imaging (MRI) plays a crucial role in clinical diagnosis and scientific research as a non-invasive imaging modality that can provide multi-contrast images. However, acquiring fully-sampled k-space data is usually time-consuming due to the physical limitations of the scanning device \cite{liang2000principles}. To address this issue, different methods have been introduced to accelerate MRI data acquisition. K-space data undersampling followed by high-quality MR image reconstruction is one of the most common techniques in the field \cite{wang2021deep}. Recently, deep learning-based methods have shown outstanding performance in accelerating MR image reconstruction. These methods can capture rich prior information from large amounts of data \cite{wang2016accelerating,hammernik2018learning, schlemper2017deep,mardani2018deep,aggarwal2018modl, zhu2018image}. Deep learning-based MR image reconstruction methods can be broadly classified into data-driven \cite{wang2016accelerating,schlemper2017deep,mardani2018deep,zhu2018image} and model-driven approaches \cite{hammernik2018learning, aggarwal2018modl}. Data-driven approaches typically rely on large amounts of data and learn mapping relationships from undersampled data to fully-sampled data. Instead, model-driven approaches unroll the traditional optimization algorithm to the network to achieve end-to-end reconstruction. Although these methods have facilitated the development of MR image reconstruction to some extent, they need to collect large amounts of data for centralized training. For some institutions, it is difficult to collect adequate data due to the expensive acquisition cost. On the other hand, aggregating data from multiple institutions raises a serious problem---data privacy leakage, which may become infeasible in a realistic healthcare scenario \cite{kaissis2021end}.

Federated learning (FL), a distributed training paradigm, allows for collaborative learning across multiple institutions while protecting privacy \cite{li2020federated,mcmahan2017communication,li2020bfederated,li2021fedbn}. The basic training processes of federated learning are as follows: 1) the server sends the initialized model to each client; 2) each client uses local computational resources and private data to train the model and sends the trained model to the server; 3) the server aggregates the models sent by the clients and broadcasts the aggregated model to the clients; 4) each client uses the aggregated model to update the local model and tests the performance of the updated model. After multiple interactive training between the clients and the server, a better global model can be obtained eventually. FedAvg \cite{mcmahan2017communication} is the most classical federated learning framework that implements aggregation by averaging the model parameters of each client. Several researchers have attempted to solve the MR image reconstruction problem using federated learning\cite{guo2021multi,feng2022specificity,elmas2022federated,gong2022federated,feng2023learning}. For example, FL-MRCM\cite{guo2021multi} is the first attempt of employing federated learning method in MR image reconstruction. It alleviates the problem of domain shift by continuously aligning the latent features of source and target clients. In addition, FedMRI\cite{feng2022specificity} achieves client-specific reconstruction by decomposing the reconstruction model into a global-shared encoder and local-personalized decoder. Although existing federated learning MR reconstruction methods have achieved promising performance, their reconstruction models are manually designed by experts, which may suffer from performance degradation when facing heterogeneous data distributions. In addition, the model parameters may be intentionally increased to improve the reconstruction performance, which undoubtedly increases the consumption of computational resources and may also result in parameter redundancy.

Neural Architecture Search (NAS) \cite{elsken2019neural} can achieve better performance with fewer computational resources through automated architecture design.
NAS methods can be roughly divided into three categories: evolutionary algorithm-based\cite{elsken2018efficient,yang2020cars}; reinforcement learning-based\cite{bello2017neural,pham2018efficient} and gradient-based methods\cite{liu2018darts,he2020milenas}. The first two methods demand heavy computational resources, while the gradient-based method can effectively save computational resources. There have been applications of NAS for MR image reconstruction\cite{yan2020neural,huang2020enhanced}. To the best of our knowledge, neural architecture search methods have not been explored for federated learning MR image reconstruction. In order to improve the performance of the federated learning MR image reconstruction model, we propose a federated neural architecture search algorithm to improve the learning ability of the model and improve the performance of MR image reconstruction. We adopt a gradient-based NAS (differentiable architecture search) algorithm, which is more efficient and requires fewer computational resources\cite{he2020milenas}. Among different gradient-based NAS algorithms, DARTS\cite{liu2018darts} is the most classical one, but its search efficiency is not high enough. Following DARTS, MiLeNAS\cite{he2020milenas} is proposed, which achieves better search performance with mixed-level optimization. Our method adopts the MiLeNAS framework, and we improve the search space according to the demand for the number of model parameters and representation ability to make it suitable for the federated MR image reconstruction task. Our main contributions can be summarized as follows: (1) To the best of our knowledge, this is the first work to study federated neural architecture search techniques for MR image reconstruction. (2) We design a new search space according to the number of parameters, feature extraction, and representation capabilities of the model to make it more suitable for the federated MR image reconstruction task. (3) We introduce an exponential moving average method into the parameter update process of the client to increase the robustness of the client model. (4) Qualitative and quantitative experimental results demonstrate the effectiveness of our method.

\section{Proposed Method}

\subsection{DL-based MR image Reconstruction}
\begin{figure}
\centering
\includegraphics[width=10cm]{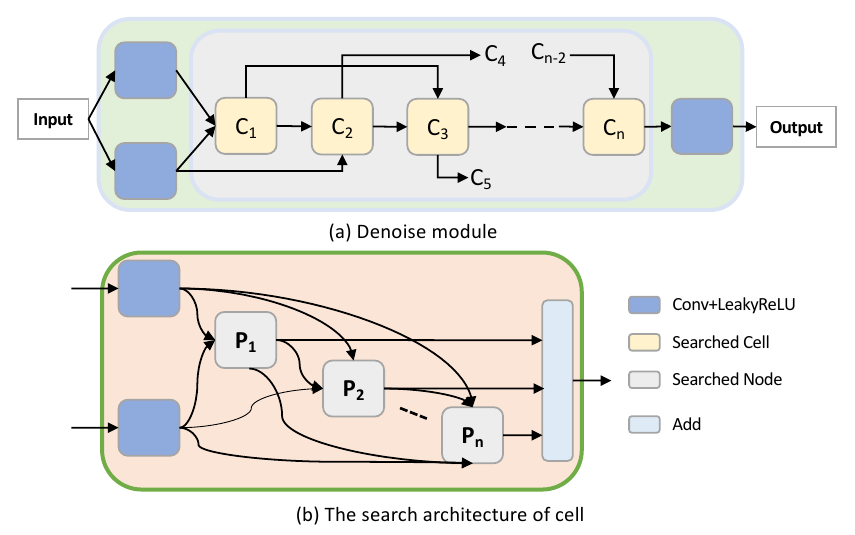}
\caption{The overall structure of search space. (a) The structure of denoiser module $D_{\omega}(\cdot)$ in our reconstruction network. The searched cells are stacked to form our basic architecture, with each cell connected to the two consecutive cells that follow it. (b) The search architecture of the cell. $P_{n}$ denotes the node in the cell.} \label{fig1}
\end{figure}

The aim of MR image reconstruction is to reconstruct a fully-sampled image $\mathbf{x} \in \mathbb{C}^{N}(M<N)$ from undersampled k-space data $\mathbf{k} \in \mathbb{C}^{M}$, such that:

\begin{equation}
\mathbf{k = Ax+\epsilon}
\label{equ1}
\end{equation}
where $\mathbf{A}\in\mathbb{C}^{M\times N}$ is the undersampling encoding matrix, and $\mathbf{\epsilon}\in\mathbb{C}^{M}$ is the measurement noise. The process of solving $\mathbf{x}$ can be transformed into the following unconstrained optimization problem:

\begin{equation}
arg\min_{\mathbf{x}}\frac{1}{2}\| \mathbf{k-Ax}\|_{2}^{2}+\lambda R(\mathbf{x})
\label{equ2}
\end{equation}
where $R(\mathbf{x})$ denotes the regularization term in the image domain, and $\lambda$ denotes the regularization coefficient. According to \cite{aggarwal2018modl}, Eq.(\ref{equ2}) can be transformed into the following alternating optimization process:

\begin{equation}
\left\{
    \begin{array}{lr}
    \mathbf{r}^{j}=D_{\omega}\left (\mathbf{x}^{j}  \right ) &  \\
    \mathbf{x}^{j+1}=\left (\mathbf{A}^{H}\mathbf{A}+\lambda \mathbf{I}  \right ) ^{-1}\left (\mathbf{A}^{H}\mathbf{k} + \lambda \mathbf{r}^{j}\right )
    \end{array}
\right.
\label{equ3}
\end{equation}
where $D_{\omega}(\cdot)$ represents a neural network for denoising,  and $\mathbf{A}^{H}$ denotes the conjugate operator of $\mathbf{A}$. The regularization parameter $\lambda$ can also be learned by the neural network. In this work, we mainly search the internal cell structure of $D_{\omega}(\cdot)$, and then stack the cell to form our denoising network. The structure is shown in Fig. \ref{fig1}.

\subsection{FedAutoMRI: FederAted neUral archiTecture search fOr MR Image reconstruction}

\subsubsection{Problem Definition.} In the federated neural architecture search setting, it is assumed that there are $C$ clients/hospitals, each of which has a private dataset $D_{c}={(x_{c}^{i},k_{c}^{i})}, i=1,..., N_{c}$. The corresponding objective function is defined as:

\begin{equation}
  arg\min_{\theta ,\alpha } \sum_{c=1}^{C} \frac{N_{c}}{N}\cdot  \{\mathbb{E}_{(x_{c},k_{c})\sim D_{c}}[\left \| f_{c}(\mathbf{A}^{H}\mathbf{k}_{c};\theta_{c},\alpha_{c})-\mathbf{x}_{c} \right \|_{2}] \}
  \label{equ4}
\end{equation}
where $\alpha_{c}$ and $\theta_{c}$ denote the architecture parameters and the corresponding model parameters of the $c^{th}$ client, respectively.

A typical NAS approach consists of three components: search space, search strategy, and performance evaluation\cite{elsken2019neural}. We designed our candidate operation set $\mathcal{O}$ based on the number of parameters and representation capabilities of the target model. The set includes (1) Standard convolution, which extracts multi-scale information through convolution kernels of different sizes: $std\_conv\_3\times3$, $std\_conv\_5\times5$ and $std\_conv\_7\times7$; (2) Dilated convolution, enlarging receptive field and reducing the number of parameters: $dil\_2\_conv\_3\times3$ and $dil\_3\_conv\_3\times3$; (3) Depthwise separable convolution, reducing parameters and computation while improving feature extraction and representation ability: $sep\_conv\_3\times3$, $sep\_conv\_5\times5$ and $sep\_conv\_7\times7$. An edge between two nodes in Fig. \ref{fig1}(b) represents one of these candidate operations. Inside the cell, to make the search space continuous by relaxing the categorical candidate operations between two nodes, mixed operations using $softmax$ over all candidate operations are performed \cite{liu2018darts}:

\begin{equation}
   \bar{\varrho}^{(j,k)}(x)=\sum_{o=1}^{|\mathcal{O} | }\frac{exp(\alpha_{o}^{(j,k)} )}{ {\textstyle \sum_{o^{'}=1}^{|\mathcal{O}|}}exp(\alpha_{o^{'}}^{(j,k)})}\varrho_{o}(x)
   \label{equ5}
\end{equation}
where $\bar{\varrho}^{(j,k)}(x)$ denotes the mixed operation for a pair of nodes $(j,k)$, $|\mathcal{O}|$ denotes the number of candidate operations, $\alpha_{o}^{(j,k)}$ denotes the weight of the $o^{th}$ operation for a pair of node $(j,k)$, and $\varrho_{o}(x)$ denotes the $o^{th}$ operation for a pair of node $(j,k)$. The aim of architecture search is to learn the encoding $\alpha=\{\alpha^{(j,k)}\}$ of the architecture.

\subsubsection{Searching Phase.}
In DARTS\cite{liu2018darts}, the architecture parameters are optimized using only the loss of the validation set, which may not be optimal. To this end, He et al., \cite{he2020milenas} proposed a mixed optimization framework MiLeNAS that exploits both training and validation losses. Specifically, Eq. (\ref{equ4}) is solved by optimizing $\omega$ and $\alpha$ during the local search process:

\begin{equation}
   \begin{cases}
\omega_{c}^{t,z+1}=\omega_{c}^{t,z}-\eta_{\omega} \bigtriangledown_{\omega}\mathcal{L}_{tr}(\mathbf{A}^{H}\mathbf{k}_{c},\mathbf{x}_{c};\omega_{c}^{t,z},\alpha_{c}^{t,z}) 
 \\
\begin{aligned}
\alpha_{c}^{t,z+1}=\alpha_{c}^{t,z}-\eta_{\alpha}\{\bigtriangledown_{\alpha}\mathcal{L}_{tr}(\mathbf{A}^{H}\mathbf{k}_{c},\mathbf{x}_{c};\omega_{c}^{t,z},\alpha_{c}^{t,z})+
\\
\beta\bigtriangledown_{\alpha}\mathcal{L}_{val}(\mathbf{A}^{H}\mathbf{k}_{c},\mathbf{x}_{c};\omega_{c}^{t,z},\alpha_{c}^{t,z})\}
\end{aligned} 
\end{cases}
\label{equ6}
\end{equation}
where $t$ and $z$ denote global communication rounds and local training epochs, respectively. $\beta$ denotes a non-negative regularization parameter that balances the importance of training and validation loss for $\alpha$. $\eta_{\omega}$ and $\eta_{\alpha}$ represent the learning rates for updating $\omega$ and $\alpha$, respectively. After the local search, all clients send the updated $\alpha$ and $\omega$ to the server, and the central server aggregates all parameters. We use a similar aggregation scheme as in \cite{he2020towards}, namely the average aggregation approach. Then, the server broadcasts the aggregated $\omega$ and $\alpha$ to all clients, and each client updates the local parameters for the next communication.  

\subsubsection{Training Phase.}
After the model architecture search, the operations with the top two weights are selected as the operations in our final model for each pair of nodes $(j,k)$. Considering the limitation of computational resources, there are only three nodes in the cell. In the training phase, in order to improve the robustness of the local client model, we introduce an exponential moving average method, which can be used to estimate the local mean of the variable. The corresponding formula is:

\begin{equation}
    \omega_{c}^{t}=\frac{\gamma \cdot \omega _{c}^{t-1}+(1-\gamma )\cdot \omega^{t}}{1-\gamma^{t}}
\end{equation}
where $\omega_{c}^{t}$ represents the weight of the $c^{th}$ client at the $t^{th}$ communication, $\omega^{t}$ represents the aggregated weight of the server at the $t^{th}$ communication, and $\gamma$ represents the weighting coefficient.


\section{Experiments and Results}
\subsection{Experimental Settings}

\subsubsection{Dataset.} We search and train our proposed framework on three public datasets. Details of the datasets are provided as follows: 1) \textbf{fastMRI}\cite{knoll2020fastmri}: A large public dataset of 1.5T and 3T data, from which we utilized T1 brain data of 1140 subjects; 2) \textbf{MoDL-Brain}\footnote{https://github.com/hkaggarwal/modl}: A total of 524 slices are provided and the matrix size is cropped to $256\times256$; 3) \textbf{CC359}\footnote{https://sites.google.com/view/calgary-campinas-dataset/mr-reconstruction-\newline challenge}: A total of 35 subjects are provided. The fastMRI dataset is used in the search phase, which is equally divided into four clients, each containing 285 objects. 

\subsubsection{Implementation Details.} All the networks were trained using the PyTorch framework with one NVIDIA RTX A6000 GPU (with 48GB memory). In the searching phase, the Adam optimizer with a learning rate of 3e-3 is used to update architecture parameters and the weight decay is set to 1e-3. Besides, the Adam optimizer with a learning rate of 1e-3 is used to update local model parameters. Networks are trained for 50 global communication rounds with 5 local epochs. In the training phase, the AdamW optimizer with the initial learning rate of 1e-3 is adopted. Networks are trained for 150 global communication rounds with 5 local epochs. 

\subsection{Comparison with State-of-the-Art Methods}
To demonstrate the effectiveness of our proposed method, we compare it with various state-of-the-art (SOTA) methods: 1) \textbf{Non-Fed}: each client is individually trained using their private data without federated learning; 2) \textbf{FedAvg}\cite{mcmahan2017communication}: a global model is obtained by using the average aggregation method; 3) \textbf{FedProx}\cite{li2020bfederated}: a global model is obtained by adding a regularization term to the loss function of the client; 4) \textbf{FedMRI}\cite{feng2022specificity}: personalized model is learned by adding a weighted contrastive regularization term to each client.

\begin{figure*}[ht]
\centering
\includegraphics[width=10cm]{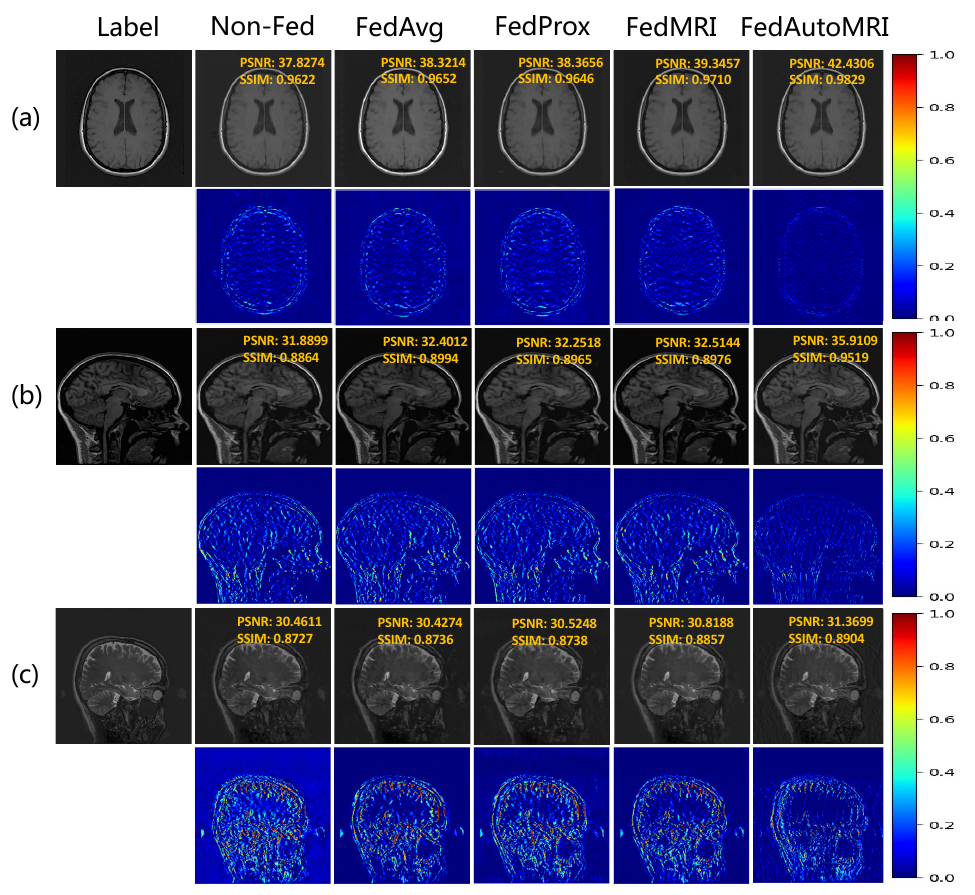}
\caption{Qualitative reconstruction results of different methods on the three datasets ((a) fastMRI, (b) CC359, (c)MoDL-Brain). From left to right, the six images corresponding to the reference image, and the reconstructed images of Non-Fed, FedAvg, FedProx, FedMRI and our FedAutoMRI, respectively. The second, fourth and six rows plot the corresponding error maps.} 
\label{fig2}
\end{figure*}

Table \ref{tab1} lists the quantitative results of different methods. Overall, our proposed FedAutoMRI achieves a better reconstruction performance when compared to the SOTA methods. Due to the small amount of data provided by the MoDL-Brain dataset, the results of the comparison methods are worse on this dataset. By introducing the exponential moving average method, the robustness of our model is improved. Thus, it obtains satisfactory performance on this dataset. In addition, the number of parameters of our model is 0.016M, much fewer than the comparison methods. In other words, our proposed FedAutoMRI achieves better performance compared to the four existing methods by using a very lightweight network architecture, which proves that our model is more efficient. In addition to the quantitative results, qualitative results are plotted in Fig. \ref{fig2}, and similar conclusions can be made that FedAutoMRI can reconstruct higher-quality MR images with smaller errors.

\begin{table}[ht]
\centering
\caption{Quantitative results of different methods on the three datasets. Bold numbers indicate the best results. }
\label{tab1}
\renewcommand{\arraystretch}{1.5} 
\resizebox{\textwidth}{!}{
\tiny 
\begin{tabular}{cccccccc}
\toprule[1pt]
\multirow{2}{*}{Method} & \multicolumn{3}{c}{PSNR/dB}                                                & \multicolumn{3}{c}{SSIM}                                               & \multirow{2}{*}{Param/M} 
\\ 
\cline{2-7}
& \multicolumn{1}{c}{fastMRI} & \multicolumn{1}{c}{CC359}   & MoDL-Brain & \multicolumn{1}{c}{fastMRI} & \multicolumn{1}{c}{CC359}  & MoDL-Brain &                   \\ \midrule[1pt]
Non-Fed                 & \multicolumn{1}{c}{35.2427} & \multicolumn{1}{c}{31.5358} & 29.7220    & \multicolumn{1}{c}{0.9488}  & \multicolumn{1}{c}{0.8990} & 0.8084     & 7.76
\\ 
FedAvg\cite{mcmahan2017communication}                  & \multicolumn{1}{c}{35.4097} & \multicolumn{1}{c}{31.6159} & 29.5923    & \multicolumn{1}{c}{0.9519}  & \multicolumn{1}{c}{0.9053} & 0.8040     & 7.76                  
\\ 
FedProx\cite{li2020bfederated}                 & \multicolumn{1}{c}{35.6159} & \multicolumn{1}{c}{31.8033} & 29.8534    & \multicolumn{1}{c}{0.9529}  & \multicolumn{1}{c}{0.9052} & 0.8148     & 7.76                  
\\ 
FedMRI\cite{feng2022specificity}                  & \multicolumn{1}{c}{36.4179} & \multicolumn{1}{c}{32.0351} & 29.8061    & \multicolumn{1}{c}{0.9594}  & \multicolumn{1}{c}{0.9105} & 0.8114     & 7.76                  
\\ 
FedAutoMRI                 & \multicolumn{1}{c}{\textbf{39.6817}} & \multicolumn{1}{c}{\textbf{35.2123}} & \textbf{31.3378}    & \multicolumn{1}{c}{\textbf{0.9782}}  & \multicolumn{1}{c}{\textbf{0.9515}} & \textbf{0.8639}     & \textbf{0.016}                 
\\ \bottomrule[1pt]
\end{tabular}
}
\end{table}

\subsection{Ablation Study}
In this section, we analyze the effectiveness of the exponential moving averaging method and the efficiency of the model obtained by searching. Table \ref{tab2} gives the corresponding results. According to these quantitative indicators, the performance of the model is improved after adding the exponential moving average, which verifies the effectiveness of this module. In addition, in order to compare the model performance with the manually designed model, we list the number of parameters and computational complexity corresponding to a single model. Compared with the baseline model, our model only needs about 6.1\% of the parameters and 5.7\% of the computation, which proves the efficiency of our searched model.

\begin{table*}[ht]
\centering
\caption{Results from ablation studies. Bold numbers indicate the best results.}
\label{tab2}
\renewcommand{\arraystretch}{1.5} 
\resizebox{\textwidth}{!}{
\tiny 
\begin{tabular}{ccccccccc}
\toprule[1pt]
\multirow{2}{*}{Method} & \multicolumn{3}{c}{PSNR/dB}                                             & \multicolumn{3}{c}{SSIM}                                               & \multirow{2}{*}{Param/M} & \multirow{2}{*}{FLOPs/G} 
\\ \cline{2-7}

& \multicolumn{1}{c}{fastMRI} & \multicolumn{1}{c}{CC359}   & MoDL-Brain & \multicolumn{1}{c}{fastMRI} & \multicolumn{1}{c}{CC359}  & MoDL-Brain &                          &                          
\\ \midrule[1pt]
baseline                & \multicolumn{1}{c}{37.0153} & \multicolumn{1}{c}{33.6756} & 30.1359    & \multicolumn{1}{c}{0.9611}  & \multicolumn{1}{c}{0.9297} & 0.8417     & 0.26                     & 85.65              \\
ema(w/o)                & \multicolumn{1}{c}{39.6064} & \multicolumn{1}{c}{35.1756} & 31.0704    & \multicolumn{1}{c}{0.9776}  & \multicolumn{1}{c}{0.9499} & 0.8546     & 0.016                    & 4.92               \\ 
ema(w)                  & \multicolumn{1}{c}{\textbf{39.6817}} & \multicolumn{1}{c}{\textbf{35.2123}} & \textbf{31.3378}    & \multicolumn{1}{c}{\textbf{0.9782}}  & \multicolumn{1}{c}{\textbf{0.9515}} & \textbf{0.8639}     & \textbf{-}                    & \textbf{-}               
\\ \bottomrule[1pt]
\end{tabular}
}
\end{table*}

\vspace{-0.5cm}
\section{Conclusions}
In this work, we proposed a federated neural architecture search framework for MR image reconstruction. We designed the search space to capture the heterogeneous data distributions, and we utilized differentiable architecture search methods to find the optimal architecture. In addition, to improve the robustness of the client model, we introduced an exponential moving average method. Experimental results validated that our method can better learn the prior knowledge from the data and obtain enhanced reconstruction performance on the three datasets. Results from ablation studies further verified the efficiency and effectiveness of our model.

\subsubsection{Acknowledgments.} This research was partly supported by the National Natural Science Foundation of China (62222118, U22A2040), Guangdong Provincial Key Laboratory of Artificial Intelligence in Medical Image Analysis and Application (2022B1212010011), Shenzhen Science and Technology Program (RCYX20210706092104034, JCYJ20220531100213029), and Key Laboratory for Magnetic Resonance and Multimodality Imaging of Guangdong Province
(2020B
1212060051).

\bibliographystyle{splncs04}
\bibliography{ref}
%




\end{document}